\documentclass[pra,showpacs,twocolumn]{revtex4}
\usepackage{bm,graphicx,amsmath}
\usepackage{amsfonts}
\usepackage{amssymb}
\usepackage{amsthm}
\begin{document}
\title{Travelling to exotic places with cavity QED systems}

\author{Jonas Larson\footnote{email: jolarson@kth.se}}

\affiliation{NORDITA, 106 91 Stockholm, Sweden}

\begin{abstract}
Recent theoretical schemes for utilizing cavity QED models as
quantum simulators are reviewed. By considering a quadrature
representation for the fields, it is shown how Jahn-Teller models,
effective Abelian or non-Abelian gauge potentials, transverse Hall
currents, and relativistic effects naturally arise in these systems.
Some of the analytical predictions are verified numerically using
realistic experimental parameters taking into account for system losses.
Thereby demonstrating their feasibility with current experimental
setups.
\end{abstract}
\maketitle

\section{Introduction}
Back in 2006 I was a co-author on a paper called {\it ``Travelling
to exotic places with ultracold atoms"} where we discussed some, at
the time, new ideas how to mimic unusual phenomena occurring in
condensed matter systems by using ultracold atoms~\cite{jonas1}. Due to
the controllability of system parameters and long decoherence times,
cold atoms in optical lattices had shown to be suitable candidates for
simulation of different many-body theories developed in the
community of condensed matter physics~\cite{maciek}. Around the same
time, many theoretical predictions were presented in which high
energy and cosmological phenomena could be simulated by means of
Bose-Einstein condensates~\cite{becsim}. In addition, ion-trap
systems have recently drawn much attention in terms of simulating
various spin chains as well as relativistic effects~\cite{ionsim}.

In the present proceeding I summarize some new directions employing
cavity QED systems as quantum simulators. On the experimental side,
cavity QED systems have seen great progress during the last years.
Pure quantum effects are nowadays not restricted to single few-level
atom-cavity setups~\cite{haroche}, but include both quantum-dots (qdots) or
SQUIDs~\cite{dotqed} as well as Bose-Einstein
condensates~\cite{becqed} coupled to single cavity modes. These new
experiments pave the way for reaching the superstrong coupling
regime where the strength of the effective matter-field coupling
exceeds the system losses by several orders of magnitude. One
crucial outcome of reaching this regime is that application of the
rotating wave approximation (RWA) becomes questionable~\cite{jonas2}. In
most cases, going beyond the RWA implies that the corresponding
system Hamiltonian is not integrable and other
approximations or numerical methods are required. As outlined in
Ref.~\cite{jonas2}, by expressing the cavity field(s) in its quadrature
operators, it directly follows that beyond the RWA typical cavity QED
models describe a set of coupled harmonic potentials. In this
representation it is thereby possible to develop new physical
intuition not easily extracted when the field is expressed
in boson creation and annihilation operators. Moreover, in this picture the link between cavity QED models and systems from other fields of physics readily follows.

We consider a single $N$-level atom~\cite{fotnot}
coupled to a set of quantized cavity modes. The quadrature operators
for mode $k$ reads
\begin{equation}
\begin{array}{lll}
\hat{X}_k=\frac{1}{\sqrt{2}}\left(\hat{a}_k+\hat{a}_k^\dagger\right),
& &
\hat{P}_k=\frac{i}{\sqrt{2}}\left(\hat{a}_k-\hat{a}_k^\dagger\right)
\end{array}
\end{equation}
with $\hat{a}_k$ ($\hat{a}_k^\dagger$) the annihilation (creation)
operator of a photon in mode $k$. The quadrature operators obey the
canonical commutation relations
$[\hat{X}_k,\hat{P}_l]=i\delta_{kl}$, and thereby can be viewed as
``position" and ``momentum" for the field. In the dipole
approximation we assume the electric field $E({\bf x})$ to be
constant over the extent of the atom, hence letting ${\bf x=0}$, to
obtain
\begin{equation}
\bar{E}=\sum_k\bar{\varepsilon}_k\mathcal{E}_k\hat{P}_k,
\end{equation}
where $\bar{\varepsilon}_k$ and $\mathcal{E}_k$ are the polarization
vector and the field amplitude of the $k$'th mode respectively.
Denoting the dipole moment between the atomic states $|i\rangle$ and
$|j\rangle$ by $\bar{d}_{ij}$, the atom-field interaction takes the
form
\begin{equation}
H_I=\sum_{i,j}\bar{d}_{ij}\cdot\bar{E}.
\end{equation}
Due to selection rules or large atom-field detunings, most terms in
the above sums vanish either exactly or approximately.
Explicitly, the dipole moment
$\bar{d}_{ij}=(d_{ij}^x,d_{ij}^y,d_{ij}^z)$, with
$d_{ij}^\alpha=-e|i\rangle\langle i |\alpha|j\rangle\langle
j|+\mathrm{H.c}$ and $e$ the electron charge. The full Hamiltonian,
containing the free field and internal atomic energies, becomes
\begin{equation}\label{ham}
H=H_f+H_a+H_I,
\end{equation}
where
\begin{equation}
\begin{array}{lll}
H_f=\hbar\displaystyle{\sum_k}\omega_k\left(\frac{\hat{P}_k^2}{2}+\frac{\hat{X}_k^2}{2}\right),
& & H_a=\displaystyle{\sum_{j=1}^NE_j|j\rangle\langle j|,}
\end{array}
\end{equation}
$\omega_k$ the frequency of mode $k$, and $E_l$ being the energy of the
$j$'th initial atomic state.

\section{Jahn-Teller systems}
Jahn-Teller systems frequently occur in both molecular/chemical
physics and condensed matter physics~\cite{jt0}. The name dates back
to the works by Hermann Jahn and Edward Teller studying symmetry
breaking for polyatomic molecules~\cite{jt}. Common for the
Jahn-Teller models is that two potential surfaces become degenerate
in a single point forming a {\it conical intersection}. About two
decades after the pioneering work by Jahn and Teller,
Longuet-Higgins showed that by encircling the degeneracy, the
electronic wave function changes sign~\cite{lh}. This phase factor
was later realized to be the so called {\it Berry
phase}~\cite{berry}.

\subsection{The $\beta\times E$ model}
\begin{figure}[h]
\centerline{\includegraphics[width=5.5cm]{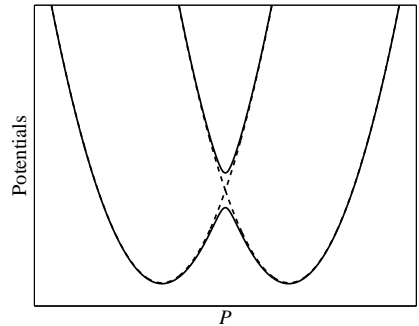}}
\caption{Typical diabatic (dashed) and adiabatic (solid) potentials
for the $\beta\times E$ model. } \label{fig1}
\end{figure}

The simplest non-trivial atom-cavity system considers a two-level
atom interacting with a single cavity mode. Introducing the Pauli
matrices $\hat{\sigma}_x=|1\rangle\langle2|+|2\rangle\langle1|$,
$\hat{\sigma}_y=i(|1\rangle\langle2|-|2\rangle\langle1|)$, and
$\hat{\sigma}_z=|1\rangle\langle1|-|2\rangle\langle2|$, the
Hamiltonian reads
\begin{equation}
H_{\beta
E}=\hbar\omega\left(\frac{\hat{P}^2}{2}+\frac{\hat{X}^2}{2}\right)+\frac{\hbar\Omega}{2}\hat{\sigma}_z+\hbar
g\hat{\sigma}_x\hat{P}.
\end{equation}
Here, $\Omega$ is the atomic transition frequency and $g$ the
effective atom-field coupling. Application of the RWA with respect
to the first two terms renders the well known Jaynes-Cummings
model~\cite{jc}. For true atomic systems, it accurately explains
several experimental observations~\cite{haroche}. In terminology of
molecular physics, $H_{\beta E}$ is equivalent to the $\beta\times
E$ Jahn-Teller Hamiltonian by simply interchanging $\hat{X}$ and
$\hat{P}$. For a deeper understanding of the dynamics, it is more
convenient to think of $\hat{X}^2/2$ as the kinetic energy term and
$\hat{P}$ as the coordinate. In this picture, the {\it adiabatic
potentials} are defined as
\begin{equation}
V_{ad}^\pm(\hat{P})=\frac{\hbar\omega\hat{P}^2}{2}\pm\hbar\sqrt{\frac{\Omega^2}{4}+g^2\hat{P}^2}.
\end{equation}
The {\it diabatic potentials} are
$V_d^\pm(\hat{P})=\hbar\omega\hat{P}^2/2\pm\hbar g\hat{P}$.
Figure~\ref{fig1} depicts examples of the different potentials.
Whenever $|g|>\sqrt{\omega\Omega}/2$, the lower adiabatic potential
$V_{ad}^-(\hat{P})$ possesses two minima, while for
$|g|<\sqrt{\omega\Omega}/2$ a single global minimum occurs at $P=0$.
The Jahn-Teller effect~\cite{jt0,jt} is easily understood from the
adiabatic potentials; the amplitude of the ground state wave
function does not attain its maximum at $P=0$, but around the two shifted minima. For polyatomic molecules it implies that the conical intersection cause the
molecule to vibrate around a less symmetric atomic configuration, or
in other words, the ground state energy is decreased by lowering the
symmetry. In terms of cavity QED, the Jahn-Teller effect states that
for sufficiently strong atom-field coupling the ground state is not
the one with the atom in its ground state and the field in vacuum.
This effect has long been known in cavity QED, and in the case of
infinitely many atoms (the thermodynamical limit) it describes a second
order quantum phase transition~\cite{dicke}. However, only recently
has it been shown that this phase transition is identified as a
Jahn-Teller symmetry breaking~\cite{jonas3}.

Before proceeding, some words of caution. The presence of a
Jahn-Teller effect in the cavity QED setting is not possible by
considering a direct coupling between the true atomic ground state
and an excited state. A full microscopic derivation of the
Hamiltonian contains a ``self-energy" term that has been neglected.
This is justified in most experimental situations, but in the
$|g|>\sqrt{\omega\Omega}/2$ regime it can no longer be left out. Once this
term is considered, sum-rules or gauge arguments can be employed to
show that the lower adiabatic potential indeed never attains two
minima~\cite{a2}. In typical experiments, however, the atomic state
$|g\rangle$ is not the true atomic ground state. Moreover, by
utilizing Raman transitions the coupling is not direct and the above
arguments do not apply. Longer discussions on this topic can be
found in Refs.~\cite{jonas3,jonas4}.

\subsection{The $\varepsilon\times E$ model}
By considering a degenerate bimodal cavity, the $\beta\times E$
model can be extended to the $\varepsilon\times E$ one provided the
atom-field couplings obey
$\bar{d}_{eg}\cdot\bar{\varepsilon}_x\mathcal{E}_x=\hbar
g\hat{\sigma}_x$ and
$\bar{d}_{eg}\cdot\bar{\varepsilon}_y\mathcal{E}_y=\hbar
g\hat{\sigma}_y$. We label the two modes by $x$ and $y$. The
Hamiltonian then takes the form
\begin{equation}\label{jtham2}
H_{\varepsilon
E}\!=\!\hbar\omega\!\!\left(\!\frac{\hat{P}_x^2\!+\!\hat{P}_y^2}{2}\!+\!\frac{\hat{X}^2\!+\!\hat{Y}^2}{2}\!\right)\!\!+\!\frac{\hbar\Omega}{2}\hat{\sigma}_z\!+\!\hbar
g\hat{P}_x\hat{\sigma}_x\!+\!g\hat{P}_y\hat{\sigma}_y\!,
\end{equation}
with the adiabatic potential surfaces
\begin{equation}
V_{ad}^\pm(\hat{P}_x,\hat{P}_y)=\hbar\omega\frac{\hat{P}_x^2+\hat{P}_y^2}{2}\pm\hbar\sqrt{\frac{\Omega^2}{4}+g^2\left(\hat{P}_x^2+\hat{P}_y^2\right)}.
\end{equation}
Instead of having the double well structure as in Fig.~\ref{fig1},
the lower adiabatic potential has a sombrero shape. This model was
extensively studied in Ref.~\cite{jonas3}, focusing on the effects
due to the non-zero Berry phase acquired when encircling the conical
intersection. Initializing a coherent state in one of the cavity
modes such that it predominantly populate the lower adiabatic
potential, it was demonstrated that over longer time periods
population with be periodically swaped between the two modes and that
this period greatly depends on the Berry phase.

\subsection{Renner-Teller model}
If the degeneracy is not conical it is glancing, meaning that the
tangents of the two potential surfaces are identical. This defines
the Renner-Teller models. The same type of Jahn-Teller effect is
possible for Renner-Teller models, but the Berry phase vanishes
whenever the intersection is in configuration space. From a physical
point of view, the model we consider is thereby different since the
intersection occurs in momentum space rather than in configurations
space.

To achieve a glancing intersection we look at a three level
$\Lambda$-atom with the two lower atomic states $|1\rangle$ and
$|2\rangle$ coupled via two cavity modes to an excited state
$|3\rangle$. The model Hamiltonian is taken as
\begin{equation}
\begin{array}{lll}
H_{RT} & = &
\displaystyle{\hbar\omega\left(\frac{\hat{P}_x^2+\hat{P}_y^2}{2}+\frac{\hat{X}^2+\hat{Y}^2}{2}\right)+\sum_{j=1,2,3}E_j|j\rangle\langle
j|}\\ & &
+g\left(|3\rangle\langle1|\hat{P}_x+|3\rangle\langle2|\hat{P}_y+\mathrm{H.c.}\right).
\end{array}
\end{equation}
As a three-level system, there are three adiabatic and diabatic
potential surfaces. Assuming degenerate modes and degenerate ground
atomic states, $E_1=E_2=0$, the adiabatic potentials become
\begin{equation}
\begin{array}{l}
V_{ad}^\pm(\hat{P}_x,\hat{P}_y)=\displaystyle{\hbar\omega\frac{\hat{P}_x^2\!+\!\hat{P}_y^2}{2}\!+\!\frac{E_3}{2}\!\pm\sqrt{\frac{E_3^2}{2}\!+g^2\!\left(\hat{P}_x^2\!+\!\hat{P}_y^2\right)}},
\\
V_{ad}^0(\hat{P}_x,\hat{P}_y)=\displaystyle{\hbar\omega\frac{\hat{P}_x^2\!+\!\hat{P}_y^2}{2}}
\end{array}
\end{equation}
from which it is clear that $V_{ad}^-(\hat{P}_x,\hat{P}_y)$ and
$V_{ad}^0(\hat{P}_x,\hat{P}_y)$ possess a glancing intersection at
the origin.

\section{Effective gauge potentials}
Gauge theories arise in a variety of fields in physics. For the most
familiar case of a charged particle in an electromagnetic field the
theory is Abelian. An example of a non-Abelian gauge theory is the
one of Yang and Mills describing strong interaction. Getting an
experimental handle of a system showing non-Abelian characteristics
is therefore very attractive and has led to many suggestions. One
possibility is adiabatically evolving systems either by means of external changes of the Hamiltonian or for ultracold atoms subjected to spatially varying light fields
\cite{jonas5}.

The general Hamiltonian (\ref{ham}) may be rewritten as
\begin{equation}
H=\hbar\sum_k\omega_k\left(\frac{\left(\hat{P}_k-\hat{A}_k\right)^2}{2}+\frac{\hat{X}_k^2}{2}\right)+\sum_jE_j|j\rangle\langle
j|+\hat{\Psi},
\end{equation}
where
\begin{equation}
\begin{array}{lll}
\hat{A}_k=-\sum_{i,j}\bar{d}_{ij}\cdot\bar{\varepsilon}_k\mathcal{E}_k/\hbar\omega_k,
& & \hat{\Phi}=-\hbar\sum_k\omega_k\hat{A}_k^2.
\end{array}
\end{equation}
The operators $\hat{A}_k$ and $\hat{\Phi}$ have the properties as
vector and scalar potential respectively. That is, they transform
appropriately under unitary transformations~\cite{jonas6}. For any
atomic basis, these gauge potentials are matrices and they are said
to be {\it Abelian} if $[\hat{A}_k,\hat{A}_l]=0$ $\forall$ $k$ and $l$
and {\it non-Abelian} for non-commuting operators. As we now
demonstrate, all three examples of the previous section render
different types of gauge potentials.

\subsection{Abelian gauge potential}
As the atom-field interaction only includes a single mode for the  $\beta\times E$ model, the vector potential consists of a single component which reads
\begin{equation}
A=-\frac{g}{\omega}\hat{\sigma}_x.
\end{equation}
Consequently, the gauge potential is Abelian.

\subsection{Non-Abelian $SU(2)$ gauge potential}
For the second model, the $\varepsilon\times E$ one, we have
\begin{equation}
(\hat{A}_x,\hat{A}_y)=-\frac{g}{\omega}(\hat{\sigma}_x,\hat{\sigma}_y)
\end{equation}
and hence $[\hat{A}_x,\hat{A}_y]=g^2\hat{\sigma}_z/\omega^2$ showing
that the gauge potential is non-Abelian.

\begin{figure}[h]
\centerline{\includegraphics[width=7cm]{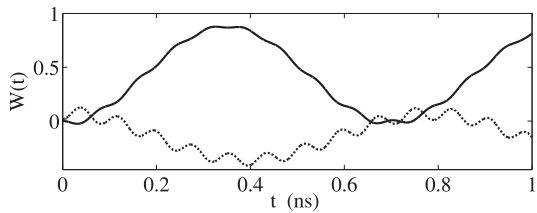}} \caption{Time
evolved atomic inversion for an initial coherent state boosted
clockwise (dotted) or anti-clockwise (solid) around the conical
intersection of the $\varepsilon\times E$ model. Parameters can be
found in Ref.~\cite{jonas6}. } \label{fig2}
\end{figure}

Naturally, the dynamics is considerably richer for a system
exhibiting non-Abelian structures. In general, time-ordering becomes
important, {\it e.g.} enclosing a ``loop" clockwise or anti-clockwise will not
result in the same final system state. In Ref.~\cite{jonas6}, the
time evolution of an initial state consisting of one empty mode and
the other with a coherent state was numerically simulated. By
properly choosing the phase of the coherent state, it will either
set off clockwise or anti-clockwise around the conical intersection.
Figure \ref{fig2} displays an example of the atomic inversion
$W(t)=\langle\hat{\sigma}_z\rangle$ during such time evolutions. The
simulation utilizes realistic parameters of the qdot cavity QED
experimented presented in~\cite{dotqed}, and furthermore takes into
account for both cavity losses and atomic spontaneous emission.
Solid and dotted lines corresponds to the different directions
around the conical intersection of the coherent state. The
difference between the two curves of the plot demonstrates the
non-Abelian property.

\subsection{Non-Abelian $SU(3)$ gauge potential}
The last of our models, the Renner-Teller one, turns out to be
non-Abelian as well. As a three-level system we express the gauge
potential in Gell-Mann matrices
\begin{equation}
(\hat{A}_x,\hat{A}_y)=-\frac{g}{\omega}(\hat{\lambda}_4,\hat{\lambda}_6),
\end{equation}
where $\hat{\lambda}_4=|3\rangle\langle1|+|1\rangle\langle3|$ and
$\hat{\lambda}_6=|3\rangle\langle2|+|2\rangle\langle3|$. We note
that $[\hat{A}_x,\hat{A}_y]=ig^2\hat{\lambda}_2/\omega^2$, with
$\hat{\lambda}_2=-i|1\rangle\langle2|+i|2\rangle\langle1|$.

\section{Anomalous Hall effect}
In the condensed matter community, the coupling of our
$\varepsilon\times E$ Hamiltonian is said to be on {\it Rashba
spin-orbit} form~\cite{so}. It is known that such coupling gives
rise to an effective Lorentz force inducing a transverse Hall
current. The force is state dependent; opposite sign for the
appropriate two internal states. When $\Omega=0$, the
population ratio between the two internal states is balanced and
therefore the Rashba coupling brings about a transverse spin
current~\cite{shall}. A non-zero $\Omega$ breaks this symmetry and
thereby it causes a net transverse particle current. Since the
phenomenon derives from an intrinsic spin-orbit coupling, and not
from an externally applied magnetic field as in the regular Hall
effect, it has been termed {\it anomalous Hall effect}~\cite{ahall}.

\begin{figure}[h]
\centerline{\includegraphics[width=5cm]{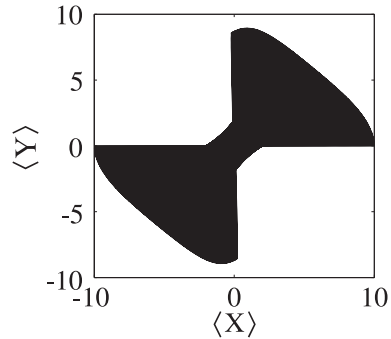}} \caption{Time
evolution of the expectations $\langle\hat{X}\rangle$ and
$\langle\hat{Y}\rangle$. Due to the transverse Hall current, the
population initially residing in mode $x$ is swaped to the $y$ mode
at the final time of propagation, which is after approximately 80 ns.
The parameters are the same as those used in Fig.~\ref{fig2}, and
can be found in~\cite{jonas7}. } \label{fig3}
\end{figure}

The net effective force acting on the state can be derived from
\begin{equation}\label{lforce}
\bar{F}=\frac{d\dot{\bar{r}}}{dt}=\frac{1}{\hbar^2}[H,[H,\bar{r}]]=g(\bar{P}\times
Z)\hat{\sigma}_z,
\end{equation}
where $\bar{r}=(\hat{X},\hat{Y})$, $\bar{P}=(\hat{P}_x,\hat{P}_y)$
is the momentum operators in the Heisenberg representation, and dot
indicates time derivative. In quantum optics, the field is
conveniently examined in phase space. The force~(\ref{lforce}) acts
on the motion of the phase space distribution. Thereby, the Hall
current occurs in phase space and hence does not involve a true
particle current. In the quadrature representation it is clear that
any field states evolving under the Hamiltonian (\ref{jtham2}) are
bounded by a harmonic potential. Without the Rashba coupling a
Gaussian coherent state will bunch back and forth in the
two-dimensional potential maintaining its chape. On the other hand,
if the Rashba coupling is non-zero, the motion of the Gaussian wave
packet will start to bend~\cite{jonas7}. Hence, the rocking motion will be
accompanied by a rotation around the $Z$-axis. This rotation is
either clockwise or anti-clockwise depending on the internal atomic state. These assumptions are verified by propagating an initial state with mode $x$ in a
coherent state with amplitude $\alpha=10/\sqrt{2}$, mode $y$ in
vacuum, and the atom in its lower state. The results are presented
in Fig.~\ref{fig3} showing the trajectories of the averages
$\langle\hat{X}\rangle$ and $\langle\hat{Y}\rangle$. It is seen that
after hundreds of oscillations in the harmonic trap, all population
has been transferred from the $x$ to the $y$ mode. The fact that the
oscillating amplitude decreases during the evolution results from
system losses.

\section{Relativistic effects}
When conical intersections as the one of the $\varepsilon\times E$
model appears in momentum space they are frequently referred to as {\it
Dirac cones}. Such Dirac cones have attracted a great amount of
interests in especially the research of graphene. In the vicinity of
the cones, the dispersions are linear as for free relativistic
particles. The electrons of graphene may therefore show relativistic
effects like for example {\it Zitterbewegung}. For avoided
intersections the relativistic electrons have a non-zero effective
mass, while for unavoided intersections the electrons are massless.

For small $\hat{P}_x$ and $\hat{P}_y$ we neglect the quadratic terms
of the $\varepsilon\times E$ Hamiltonian
\begin{equation}
H_{rel}=\hbar
g\sum_{k=x,y}\hat{\sigma_k}\hat{P}_k+\hbar\frac{\Omega}{2}\hat{\sigma}_z+V(\hat{X},\hat{Y}).
\end{equation}
This has the form of a Dirac equation for a spin-less particle
moving within a two-dimensional harmonic potential
$V(\hat{X},\hat{Y})$~\cite{dirac}. The effective masses are equal
but with opposite signs for the positive and negative energy
solutions. In the present model, small values of $\hat{P}_x$ and
$\hat{P}_y$ imply that the field amplitudes should be small,
something easily achieved with ultracold high-$Q$ cavities.

One interesting observation presented in~\cite{dosc} is that in the
non-relativistic limit, the {\it Dirac oscillator} in 2+1 dimensions
\begin{equation}
H_{Do}=c\mathbf{\alpha}\cdot(\mathbf{p}-im\beta\omega\mathbf{r})+\beta
m c^2,
\end{equation}
with $c$ the speed of light, $\mathbf{p}$ and $\mathbf{r}$ momentum
and position respectively,
$\alpha_k=\mathrm{off-diag}(\hat{\sigma}_k,\hat{\sigma}_k)$, and
$\beta=\mathrm{diag}(1,-1)$, becomes identical to an
$\varepsilon\times E$ Jahn-Teller Hamiltonian. Somewhat surprising
is that the spin-less version of $H_{Do}$, in which the Dirac
four-component matrices $\alpha_k\rightarrow\hat{\sigma}_k$ and
$\beta\rightarrow\hat{\sigma}_z$, can be mapped onto the
Jaynes-Cummings model~\cite{solano}. Furthermore, even in the
non-relativistic limit the trembling "motion" characterizing {\it
Zitterbewegung} survives and is interpreted as the Ramsey
interferometric effect~\cite{ramsey}.

\section{Conclusions}
In this paper we have given a short summary how different cavity QED
settings may serve as quantum simulators in various fields of
physics. By working in a quadrature representation for the fields,
the dynamics of the combined atom-field system can be thought of as
an artificial particle moving on a set of coupled potential
surfaces. From thereon it is easy to identify different model
Hamiltonians as Jahn-Teller ones. By rewriting these Hamiltonians, we
defined effective gauge potentials, both Abelian and non-Abelian.
Finally, we also demonstrated that relativistic effects should
appear for weak field amplitudes.

From the list of references it is clear that it is only recently
that these systems has been considered as quantum simulators. A
natural conclusion thereby is that much more is to be discovered
within this topic. To mention a few possibilities; spin Hall effects
in for example the Renner-Teller model, Dicke models with multiple
number of atoms, relativistic effects, spintronics.

This work did not cover a current proposal for simulation of
Lipkin-Meshkov-Glick many-body model by means of cavity
QED~\cite{many}.

\begin{acknowledgments}
The author acknowledges support from the MEC program (FIS2005-04627)
and Stig Stenholm for encouraging supervision during my PhD.
\end{acknowledgments}


\begin{thebibliography}{999}
\bibitem{fotnot} We will use the word ``atom", but it is understood that it may 
also represent a qdot.
\end{thebibliography}
\end{document}